# Comparison of different bonding techniques for efficient strain transfer using piezoelectric actuators

**Dorian Ziss[a]\***, **Javier Martín-Sánchez[a]**, **Thomas Lettner[b]**, **Alma Halilovic[a]**, **Giovanna Trevisi[c]**, **Rinaldo Trotta[a]**, **Armando Rastelli[a]** and **Julian Stangl[a]**

[a]Institute of Semiconductor and Solid State Physics, Johannes Kepler University, Altenbergerstraße 69, Linz, 4040, Austria

[b]Royal Institute of Technology, KTH, Brinellvägen 8, Stockholm, SE-100 44, Sweden

[c]IMEM - CNR Institute, Parco Area delle Scienze 37/a, Parma, 43124, Italy

Correspondence email: dorian.ziss@jku.at

**Abstract** In this paper strain transfer efficiencies from single crystalline piezoelectric lead magnesium niobate-lead titanate (PMN-PT) substrate to a GaAs semiconductor membrane bonded on top are investigated using state-of-the-art x-ray diffraction (XRD) techniques and finite-element-method (FEM) simulations. Two different bonding techniques are studied, namely gold-thermo-compression and polymer-based SU8 bonding. Our results show a much higher strain-transfer for the "soft" SU8 bonding in comparison to the "hard" bonding via gold-thermo-compression. A comparison between the XRD results and FEM simulations allows to explain this unexpected result with the presence of complex interface structures between the different layers.

## 1. Introduction

Piezoelectric materials have gained importance in terms of reversibly transferring strain to other materials, especially semiconductors, for the purpose of tuning the electrical and optical properties (Shayegan *et al.*, 2003; Seidl *et al.*, 2006).

The magnitude of strain induced in this way is directly proportional to the electric field applied across the piezoelectric material, making use of the converse piezoelectric effect (Kholkin *et al.*, 2008). In the past several years, devices using this approach have been extensively studied to tune the optical emission spectra of direct band-gap semiconductors (Trotta *et al.*, 2012; Huo *et al.*, 2013; Zhang *et al.*, 2013; Plumhof *et al.*, 2013; Trotta *et al.*, 2015; Martín-Sánchez *et al.*, 2016; Seidl *et al.*, 2006; Kremer *et al.*, 2014). The design of these devices has become more and more refined, which finally allows the choice of the preferred orientation of the transferred strain (Trotta *et al.*, 2015; Trotta *et al.*, 2016; Martín-Sánchez *et al.*, 2016). The techniques used to bond the semiconductor on the piezoelectric substrate are mostly gold-thermo-compression or bonding mediated by a comparatively "soft" polymer.

One of the major issues for an efficient strain transfer between the piezoelectric substrate and the semiconductor membrane seems to be the stiffness of the bonding interlayer which is mainly determined by the material's Young's modulus. In this regard, a good choice is to bond the membrane with a gold interlayer by gold-thermo-compression bonding (Kurman & Mita, 1992). However, relatively high mechanical pressures of about 10 MPa are usually required to ensure a good bonding quality. This is especially relevant when working with fragile substrates or complex device layouts where mechanical pressures in the order of only 1 kPa would be desirable (see (Trotta *et al.*, 2015)). A convenient "soft" approach is based on polymer-based SU8 bonding with a relatively high Young modulus in the order of about 5 GPa when thermally treated at temperatures above 180 °C. In addition, the SU8 polymer is commonly used as a photoresist for many processes involving lithography and bonding (LaBianca & Gelorme, 1995; Cremers *et al.*, 2001; Nallani *et al.*, 2003; Yu *et al.*, 2006).

Here, we investigate the strain transfer capabilities of gold and SU8 bonding interlayers by x-ray diffraction measurements on 330-nm-thick GaAs membranes bonded on single crystalline PMN-PT piezo-actuators. The strain transfer is analyzed by simultaneously acquiring the XRD reciprocal space maps of the bonded semiconductor film and the underlying single crystalline PMN-PT for different electric fields applied to the PMN-PT substrate. The results presented in this work provide in-depth understanding of the assets and weaknesses of the gold-thermo-compression and SU8 bonding techniques for an optimized exploitation of hybrid semiconductor-piezoelectric devices.

## 2. Sample layout and fabrication

The investigated devices are composed of a stack of three individual parts and share a common layout. The first layer is a gold-plated AlN chip carrier for electrically contacting the device, followed by the piezoelectric actuator consisting of a 225-μm-thick PMN-PT substrate and, as third layer, the 330 nm thick GaAs membrane on top; cross-sections of the devices can be seen in Fig. 1(a) and (b), respectively.

The piezoelectric material PMN-PT with the composition of $[Pb(Mg_{1/3}Nb_{2/3})O_3]_{0.71}$-$[PbTiO_3]_{0.29}$ is used due to its high piezoelectric coupling coefficients, compared to other commonly used ferroelectrics, such as lead-zirconate-titanate (PZT), and lead-zirconate-niobate (PZN) (Park & Shrout, 1997; Jiang *et al.*, 2003). It also presents good piezoelectric response at cryogenic temperatures which is very interesting for fundamental studies (Herklotz *et al.*, 2010; Bukhari *et al.*, 2014). The single crystalline PMN-PT in general has a perovskite structure (Slodczyk & Colomban, 2010) and in the composition that was used, x-ray diffraction measurements showed that the material was in a monoclinic phase (Singh *et al.*, 2006) with measured lattice constants of $a = b = 5.68 \pm 0.01$ Å and $c = 4.03 \pm 0.01$Å.

The first fabrication step is the metallization of the PMN-PT substrate on both sides by depositing a [Cr (10 nm) - Au (100 nm)] bi-layer for electrical contacting. The membrane is part of a multilayer structure [GaAs[001] substrate - $Al_{0.7}Ga_{0.3}As$(50nm) - GaAs (330nm) membrane], which is epitaxially grown by molecular beam epitaxy (MBE) and then coated by thermal evaporation with the same [Cr (10 nm) - Au (100 nm)] bi-layer. This layer serves both as a bonding layer and for protection of the membrane during the processing. In addition, the metallization of the semiconductor layer increases the spectral emission efficiency by acting as a mirror, which is important for optical measurements.

The next step is the bonding procedure, where the multi-layer structure containing the GaAs membrane is bonded on the PMN-PT substrate via gold-thermo-compression or SU8 bonding. In the case of gold-thermo-compression bonding, the semiconductor layer and the piezoelectric substrate are pressed together with a pressure of ~10 MPa and is simultaneously heated up to 300 °C for 30 minutes to enhance inter-diffusion of the gold layers and thus to form a uniform bonding layer between both parts. For the SU8 bonding, the semiconductor sample is coated with a 500 nm thick SU8 polymer by spin-coating and baked for 5 minutes at 90 °C to evaporate solvents present in as-spinned SU8. Then, the semiconductor is pressed against the PMN-PT substrate by applying a comparatively small mechanical pressure of about 10 kPa while keeping a temperature of about 220 °C for 15 minutes. This step is conducted above the glass-transition-temperature (Feng & Farris, 2002) and hardens the SU8 for an efficient bonding. It should be mentioned that a void-free bonding layer is expected when using SU8, as all possible gaps between the gold-coated piezoelectric- and semiconductor-layers are filled during the bonding process, while the SU8 is still "liquid" before the final baking step.

After bonding, the GaAs membrane is released from the grown multilayer structure onto the PMN-PT substrate by back-etching. This process consists of three steps: i) rough non-selective chemical etching of most of the GaAs substrate with $H_3PO_4 : H_2O_2$ (7:3); ii) removal of the remaining GaAs substrate down to the $Al_{0.7}Ga_{0.3}As$ sacrificial layer by selective etching with citric acid: $H_2O_2$ (4:1); iii) etching of the $Al_{0.7}Ga_{0.3}As$ layer by dipping in hydrofluoric acid (HF - 49%). More details on the device fabrication can be found in (Martín-Sánchez *et al.*, 2016).

Finally, the PMN-PT with the bonded GaAs membrane on top was glued with silver paint onto an AlN chip carrier (final sample see Fig. 1(c)). Cross sections of the differently bonded samples are shown in Fig. 1(a) and 1(b). To operate the piezoelectric device, it is necessary first to pole the piezoelectric substrate properly. This is done by applying a voltage on top of the substrate progressively in steps of 1 V up to a total voltage of 150 V. This leads to a ferroelectric ordering of the polarization in the PMN-PT domains. After poling, the applied electric field can either be parallel to the net-polarization of the piezoelectric domains or it can point in the opposite direction (anti-parallel). Since the piezo actuator is processed in such a way that it is working in the longitudinal extension mode along the [001], respectively the *z*-direction, an electric field parallel to the net-polarization results in an expansion perpendicular to the surface and an in-plane contraction. Applying the electric field in the opposite direction results in a contraction along the *z*-direction and an in-plane expansion, as long as the coercive field (~2 kV/cm for the used material) is not exceeded. To avoid such unintentional re-poling, all experiments were carried out in the first configuration with the electric field parallel to the polarization direction, i.e., inducing compressive in-plane strains.

## 3. X-ray diffraction measurements

XRD measurements were performed with a semi-commercial setup consisting of a rotating anode (Bruker AXS) in combination with XENOCS mirror optics to collimate the beam vertically and horizontally, followed by a Ge[220] channel-cut crystal monochromator. The sample was mounted on a 6-axis diffractometer (manufactured by Huber diffraction GmbH) with the possibility to apply high voltage to the sample during XRD measurements. All experiments were performed in co-planar scattering geometry with a high incidence and low exit angle to achieve a small beam footprint on the sample. The CuK$\alpha_1$ line was used with the corresponding wavelength of $\lambda=1.5406$Å. A final shaping of the beam was achieved by using adjustable slits, resulting in a beam size of 0.5×0.5 mm$^2$.

Reciprocal space maps (RSMs) around the [004] and [224] Bragg peaks of GaAs (on-top) and the [002] and [113] Bragg peaks of PMN-PT underneath were recorded using a position sensitive detector (Bruker VANTEC 1). Details on reciprocal space and XRD can be found in (Pietsch *et al.*, 2004). All reflections were recorded without changing the geometry or moving the sample in the beam. This offered the advantage that a direct comparison of the strain induced in the PMN-PT and on the corresponding area on top, in the GaAs membrane, could be measured simultaneously. For each voltage (corresponding to a certain electric field applied across the piezo) RSMs of the GaAs and PMN-PT reflections were recorded. Then, the voltage was increased in a range between 0 V and 200 V with a step size of 25 V. After each voltage ramp (e.g. 0 V-25 V @ 1 V/sec) there was a break of about 20 minutes, which was needed to limit drifting effects of the PMN-PT (Ivan *et al.*, 2011) that could result in a blurring of the Bragg-peaks.

To extract the lattice parameters and hence the strain in each material (GaAs or PMN-PT), the first step was a tilt correction applied to all RSMs by shifting the symmetric peak positions to the $\omega = 2\theta/2$ condition. For this purpose, the center-of-mass (COM) position ($\vec{Q}_{COM}$) from the symmetric RSMs [004] and [002] was calculated, as described by the following equation:

$$\vec{Q}_{COM} = \frac{1}{I_{tot}}\sum_i I_i \vec{Q}_i$$

$I_{tot}$ is the total integrated intensity of the RSM, the vector $\vec{Q}_i$ is one defined position in the RSM with the corresponding intensity $I_i$ and the index $i$ counting all measured positions in the q-space. The angle between the calculated $\vec{Q}_{COM}$ position for the sym. RSMs and the crystalline direction ($\vec{Q}_\parallel$ [001]) was used as a ω-offset. This calculated offset was also applied to the asymmetric [224] and [113] RSMs, respectively. After the tilt correction, a second COM calculation was performed for the asymmetric RSMs [224] and [113] to find the Q-in-plane component, which was finally used for evaluation of the in-plane lattice parameters and the corresponding in-plain strain values. Fig. 2 shows RSMs of GaAs and PMN-PT and the corresponding calculated $\vec{Q}_{COM}$ positions for different voltages applied.

Due to inhomogeneities in terms of pre-strains (see (Martín-Sánchez *et al.*, 2016)) and tilts after bonding (which can be seen in the RSMs as peak-broadening, distortions or side maxima), using the COM calculation was necessary to reproducibly track the parts in the RSMs which are related to *strain changes* upon bias variations. This is especially true for the RSMs of the thin GaAs membrane on-top, where most of the inhomogeneities are induced during the bonding procedure. The peak-broadening in the RSMs measured for PMN-PT is attributed to the presence of multiple domains with a small but finite angular orientation distribution ("mosaicity") in the order of 0.2° within the illuminated area(see Fig. 2). Peak-width effects are the main contribution to the error of the measured strain component. Atomic force images (AFM) taken from the polished PMN-PT surface are discussed in detail in the next section and confirm the presence of these domains.

## 4. Discussion of results and simulations

Summarizing the results from the XRD measurements, in Fig. 3 the changes of the in-plane strain component for both bonding techniques and different electric fields applied can be seen. The strain changes in the bonded GaAs membrane are for both bonding techniques *lower* than the ones measured in the PMN-PT actuator which indicates that the strain transfer is not perfect, i.e., the ratio of the strain changes is smaller than 100%. By calculating a linear regression for each set of data-points, a characteristic slope (Δε/F) can be obtained which makes it possible to quantify the strain transfer rates. The calculated slopes with their corresponding errors and the transfer efficiencies (strain induced in the PMN-PT actuator equals 100%) are given in Table 1. Interestingly, the sample bonded via the SU8 coating shows a much *high*er transfer efficiency (69%) than for gold thermo-compression bonding (25%) although the Young's modulus of hardened SU8 (≈2-4GPa (Gao et al., 2010)) is about 20 times *lower* than the modulus of a thin gold layer (≈60-70Gpa (Birleanu et al., 2016)). This seems counterintuitive in the first place, as one might think that a harder layer results in a higher strain transfer, while the opposite is observed here. The bonding efficiency is actually very sensitive to the interface properties which will be discussed below and just looking at the material parameters is not sufficient to understand and model the strain transfer correctly. We note that an almost complete strain-transfer was previously reported for epoxy-glued samples in (Shayegan *et al.*, 2003), although the reached strain levels were about an order of magnitude lower than those achieved here.

FEM simulations for different materials used as bonding layers and various interface-structures were performed to allow a deeper understanding of the strain losses. The first step was to transfer the device to an idealized model by rebuilding each individual layer using the appropriate elastic material constants and preserving the original length scales of the device. The material parameters were taken for: PMN-PT from (Luo *et al.*, 2008); the polymer SU8 from (Feng & Farris, 2002); the thin gold layers from (Birleanu *et al.*, 2016) and the GaAs layer from (Levinshtein *et al.*, 1996). All materials used were assumed to be linear-elastic.

Interestingly, the choice of the material mediating the bonding process (gold or SU8), had no significant influence on the simulated strain transferred from the piezo to the semiconductor, assuming perfectly bonded interfaces. The strain transfer is always 100%. Even for much softer hypothetic bonding materials such as rubber-like silicone polymers (Lötters *et al.*, 1997) (Young's Moduli of about two orders of magnitude lower than the Modulus of SU8), no significant strain losses could be observed in the FEM simulations. This, on first sight counter-intuitive, behaviour can be understood considering the dimensions of the structure in terms of length scales. If the layer-thickness is much smaller than the lateral dimensions of the structure (which is the case here, since layer thicknesses are in the range of $10^{-7}$ m while lateral dimensions are about $10^{-3}$m) the elastic strain induced and transferred by the individual layers has no possibility to relax except at the edge regions of the structure. Hence, only at the edges (on length scales similar to the top layer thicknesses) significant strain losses are observed due to elastic relaxations, whereas in the sample center the strain is transferred without any losses from the piezo carrier to the semiconductor membrane, regardless of the materials used for bonding. Strain relaxation is thus relevant only close to structure edges (Zander *et al.*, 2009) or for structures with high aspect ratio (Kremer *et al.*, 2014). This is of course true only within the elastic limit, i.e., if no plastic relaxations or crack formations occur. For Au bonding, this should be the case for the material constants and strain ranges in the order of 0.1%. For SU8, we will discuss this limit below.

In Fig. 4 simulations of the strain transfer efficiency (TE) for the in-plane strain component along the surface can be seen. The edge effect is clearly visible, whereas the bulk is strained uniformly. The strain transfer efficiency is quantified by color-coding the relative difference of a strain component ($\varepsilon_{yy}$) in the GaAs and PMN-PT which is given by:

$$\text{TE} = \frac{\varepsilon_{yy}(Simulation)}{\varepsilon_{yy}(PMN-PT)}$$

$\varepsilon_{yy}(Simulation)$ is the simulated strain value and $\varepsilon_{yy}(PMN-PT)$ the strain induced in the PMN-PT substrate. For the maximum of 100% strain transfer, $\varepsilon_{yy}(Simulation)$ equals $\varepsilon_{yy}(PMN-PT)$.

In contrast to the simulations, the measured strain losses for the bulk are considerable and can be explained by a more complicated interface structure. For this purpose, AFM measurements of the PMN-PT surface before and after gold coating were performed and revealed a domain-like structure even though the PMN-PT is purchased as single crystalline material. This domain-like structure appears after the surface of the piezo material is polished to reduce the initial roughness and is also responsible for the mosaicity observed in XRD. The metallization of the piezo crystal does not bury or smoothen the observed domains, it conformally reproduces the PMN-PT surface texture on the surface of the 100-nm-thick gold layer, as shown in Fig. 5. The lateral size of a single domain is in the range of 1μm with a peak-to-valley height of about 4-6 nm. This could lead to void areas, when the two Au-coated surfaces are brought into contact during the bonding process, resulting in a bonding surface area significantly below 100%. SEM images of cross-sections through gold bonded devices fabricated in the same way can be found in (Trotta *et al.*, 2012) and clearly confirm the presence of void areas where no bonding could be established. Thus, qualitative simulations on imperfect bonding interfaces were performed to explore the effect of strain losses due to void areas. In Fig. 6 simulations of a rough surface and its effect on the transfer efficiency can be seen. The rough surfaces have been qualitatively reproduced replacing the perfect bonding-layers by a regular pattern of truncated pyramids on both bonding faces. The top areas of the truncated pyramids from the gold-coated piezo and the mirrored ones from the gold-coated semiconductor are assumed to be in perfect contact, whereas the area between the truncated pyramids represents the voids. By changing the size of the top facets of the truncated pyramids, it is possible to continuously simulate a global bonding ratio between 100% (perfect bonding, whole area in contact) and 0% (no bonding established). The void parts or inhomogeneities introduced with the pyramid-like surface pattern creates "micro-edge" effects on every imperfect bonding domain and allows a partial relaxation of the induced strain. That is why the presence of defects or inhomogeneities is crucial for the relaxation of strain and hence to explain the losses in transferred strain. The effect of bonding inhomogeneities on the transferred strain can be seen in Fig.7, which reveals that the transfer efficiencies are directly correlated to the bonded area.

Hence in the case of gold-bonding the losses in strain can be very well explained by a reduction of the effective bonding area due to the intrinsic domain structure of the piezo carrier and a possible additional roughness induced during gold deposition. The efficiency of strain transfer for a particular sample depends, however, on the particular details of the bonding surfaces and cannot easily be predicted quantitatively.

For the SU8 bonded samples, the domain structure of the piezo substrate should not have any influence because the liquid SU8 could compensate the surface roughness by filling up all gaps, qualitatively explaining a higher strain transfer. However, also for SU8 bonding, the strain transfer is significantly below 100%, i.e. also in this case, the bonding layer cannot be homogeneous and continuous. Taking a closer look at the stress components in the bonding layer for measured values of strain induced in the piezo material, one can see that the elastic limit of the SU8 (65-100MPa (Spratley *et al.*, 2007)) is exceeded at the sample edges or "defects" in the interface. We believe that any such defects or inhomogeneities must trigger plastic deformation or the formation of small cracks, i.e., lead to a certain degree of plastic relaxation. This most probably occurs already during the first poling of the device, which needs to be done after the high temperature bonding step where the Curie-temperature $T_C \approx 127°$ of PMN-PT (Herklotz *et al.*, 2010) is exceeded. Afterwards, the partly plastically relaxed SU8 layer behaves elastically. I.e., during cycling the applied voltage several times over the full range the same strain state is reached for the same applied bias reproducibly. A direct confirmation of this assumption, e.g. by SEM inspection of cross-section specimen is difficult since the interfaces are hard to access while the devices are working. Further investigations are currently in progress.

**5. Conclusion**

X-ray diffraction measurements clearly show that the devices fabricated with SU8 as bonding layer show a superior efficiency in terms of strain transfer compared to the devices fabricated with gold-thermo-compression. Considering the different material constants, the measurements seem to be counter-intuitive: a "softer" bonding layer leads to higher transferred strains. Furthermore it should be mentioned that the measured strain transfer rates of $\approx 70\%$ for the devices with the "soft" SU8 bonding layer are even higher than the transfer rates reported for semiconductor-layers epitaxially grown on PMN-PT substrates (17% (Bai *et al.*, 2014); 40% (Heo *et al.*, 2016)).

Simulations on differently modelled bonding-layer-surfaces revealed that the interface structure is actually more important than the material parameters of the bonding layer. These imperfect interface structures can explain the measured losses in the transferred strain even if detailed quantitative simulations are not possible due to the interface complexity. Nevertheless, these simulations still allow a deeper understanding of processes involved during the bonding and reveal the reasons for losses in transferred strain.

Each of the studied devices is, of course, individually fabricated and simplified simulation on these devices cannot fully predict their behaviour. Therefore, for determining the exact amount of strain transferred, direct strain measurements via independent methods such as x-ray diffraction are obligatory.

**Acknowledgements**    For valuable discussions special thanks goes to Elisabeth Lausecker and Marc Watzinger. For the technical support the authors want to thank Ursula Kainz, Stephan Bräuer and Albin Schwarz. The work was supported financially by the European Union Seventh Framework Program 209 (FP7/2007-2013) under Grant Agreement No. 601126 210 (HANAS), the AWS Austria Wirtschaftsservice, PRIZE Programme, under Grant No. P1308457, by the Austrian Science Fund (FWF), under Grant No. P29603, as well as by the ERC Starting Grant No. 679183 (SPQRel).

FIG. 1. Schematic cross sections of both sample designs and the final layout of the processed devices. The numbers in brackets correspond to the layer thicknesses. (a)-(b) Schematic the cross-section of the gold and SU8 bonded sample, the gold layers mediating the bonding (see (a)) are marked with arrows inside. (c) The final layout of the processed device depicted from an angle above. The dark grey area on top is the GaAs membrane followed by the gold coated PMN-PT which is all together mounted on a chip carrier (light grey) and connected with thin aluminium wires to conductive pads where the voltage is applied (indicated by the plus and minus signs).

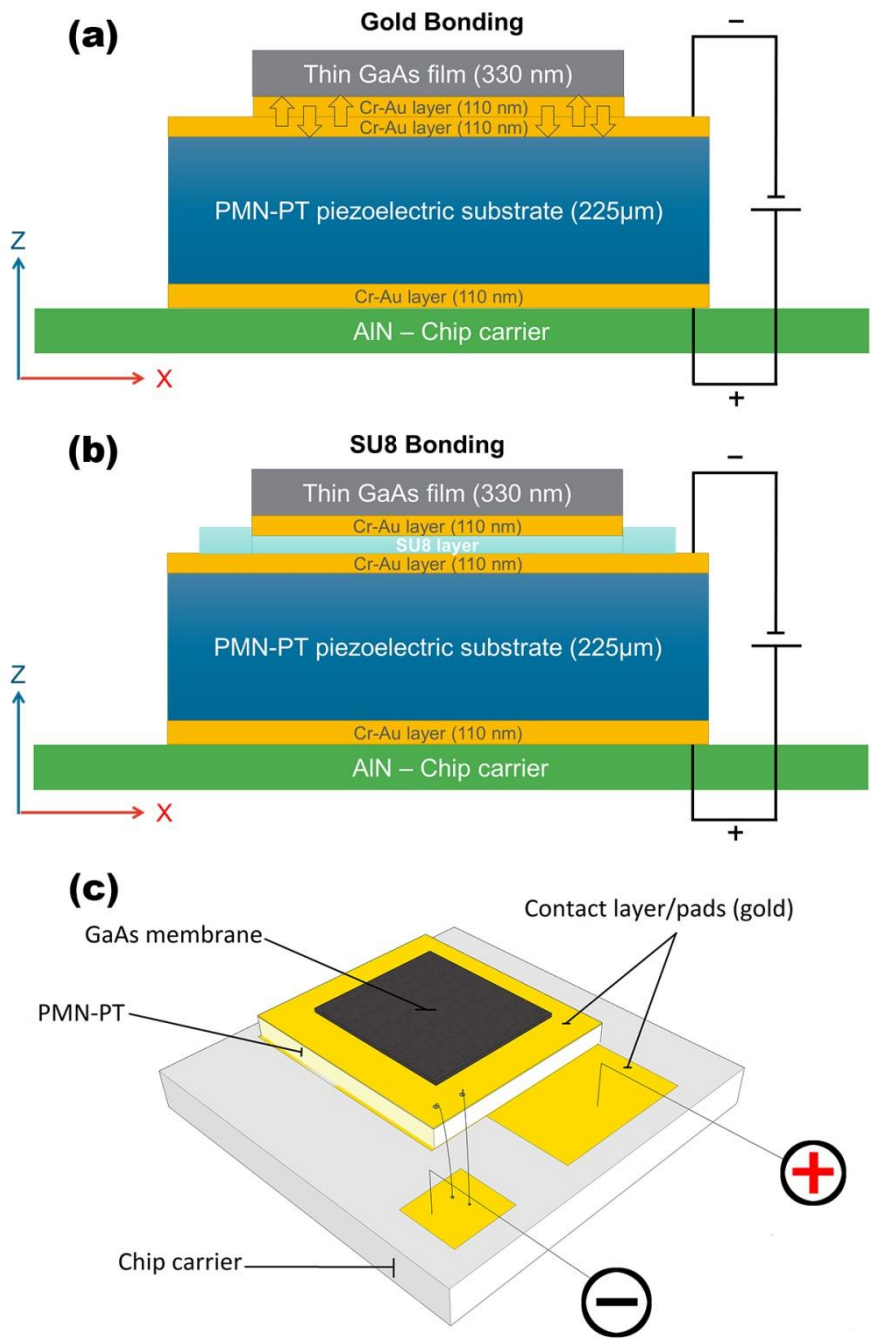

FIG. 2. RSM of GaAs [224] and PMN-PT [113] for 0V and 200V applied to the piezo actuator. The white mark shows the centre-of-mass position, $\vec{Q}_{COM}$, calculated for each RSM. In each RSM the broadening of the peaks caused by strain and tilt variations is clearly visible.

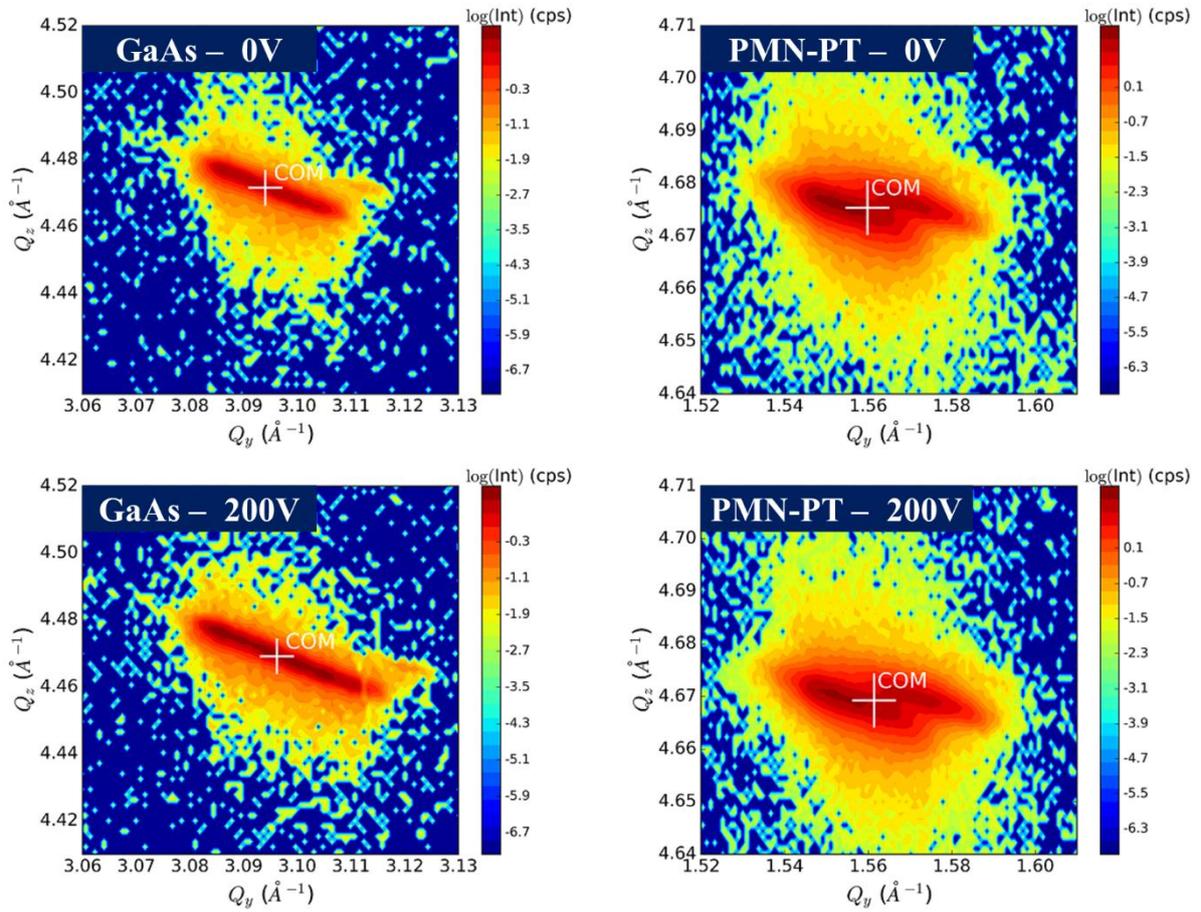

FIG. 3. Changes of the in-plane strain component ($\varepsilon_\parallel$) for the PMN-PT actuator and the GaAs membrane versus electric field, for gold bonding (a) and SU8 bonding (b). The areas marked in red represent the loss in strain. The error for the given strain values is around 0.02% and is mainly dominated by the finite peak widths in the RSMs due to mosaicity in the piezo actuator and the bonded GaAs.

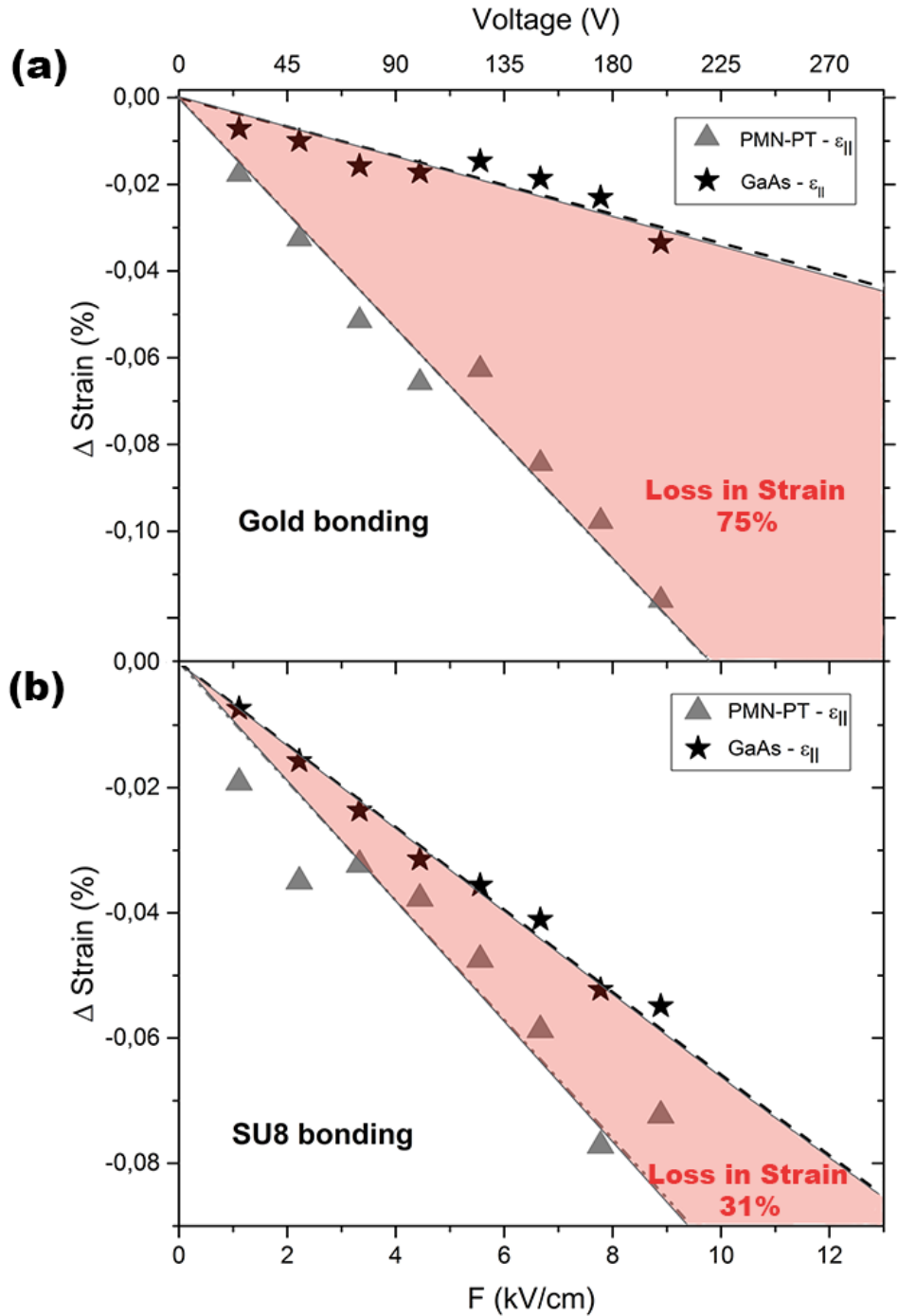

FIG. 4. Simulations of the in-plane strain ($\varepsilon_{yy}$) transfer efficiency (TE) for the SU8 bonded device. It is clearly visible that the only losses in transferred strain from the PMN-PT piezo carrier to the GaAs membrane on-top are observed in the edge regions. The effect of strain loss is observed for all edges but only the strain component ($\varepsilon_{yy}$) along the y-axis is plotted which explains the asymmetry. For a perfectly bonded device, as shown, the observed losses along the edges are only a few tenth of a percent, see colour scale. The piezo actuator was set to a compressive strain of -0.1%, which was the maximum measured strain (at 200V applied bias).

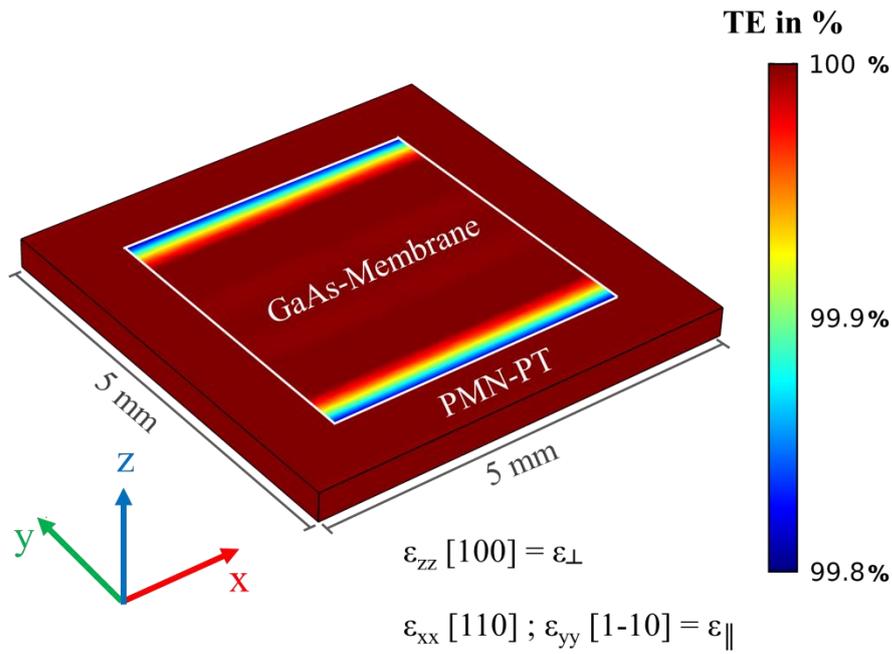

FIG. 5. AFM image of the PMN-PT surface after Au metallization. The domain-like structure can be clearly seen. The inset shows a height profile along a 5µm line along the surface and allows an estimation of the valley-to-peak distances between the individual domains which is about 4-6 nm.

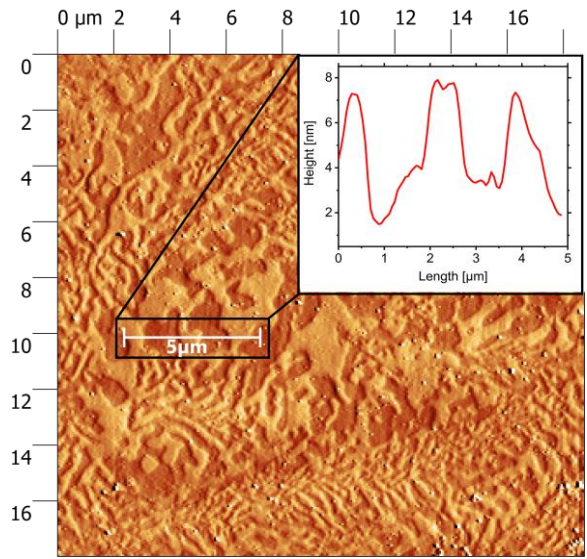

FIG. 6. Simulations of the in-plane strain ($\varepsilon_{yy}$) transfer efficiency (TE) for a patterned bonding layer with an array of truncated pyramids mimicking a rough surface. In the simulation shown in the left panel, only 10% of the total area contributes to the bonding. The size of one bonding element in this simulation is about 1µm x 1µm, which is in the range of the measured domain-sizes; the total lateral model size was reduced due to limited computing power. The reduction of the lateral dimensions lead, as expected, to very prominent edge effects visible along the y-direction for the simulated strain component $\varepsilon_{yy}$. Nevertheless, also in the centre of the structures an overall reduction of the transferred in-plane strain can be observed.

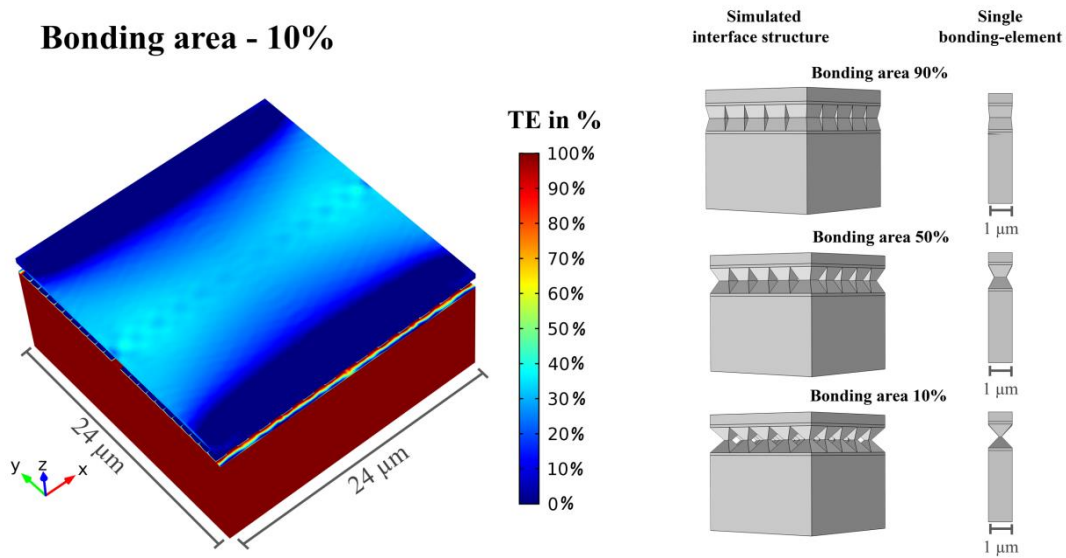

FIG. 7. Average strain transfer obtained from a volume-integration in the GaAs layer over the region (12 μm x 12 μm) marked in the inset which is not affected by the edge effect. The plot is normalized to 100% strain transfer at 100% bonding area to further eliminate any remainder of edge relaxation effects. The plot confirms that a reduction of the bonding area due to domain-like bonding elements leads to substantial losses in transferred strain.

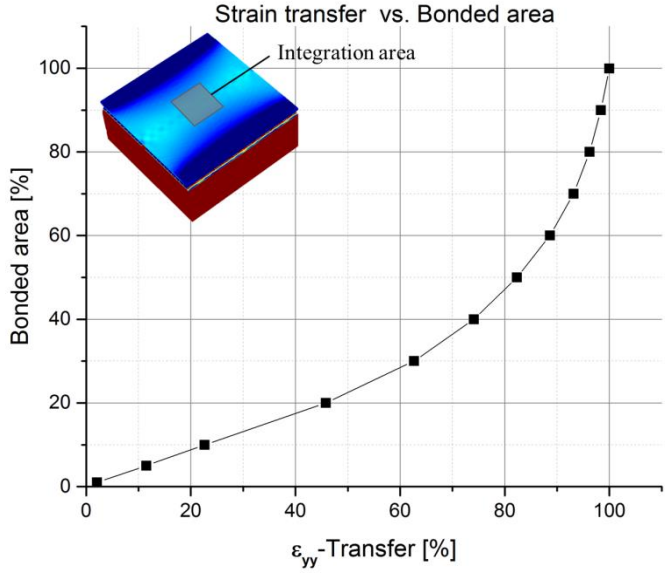

TAB. 1. Strain transfer efficiency $\Delta\varepsilon/F$ determined from XRD for the two different bonding techniques, gold thermo-compression and bonding mediated by the polymer SU8.

| Sample | PMN-PT - $\Delta\varepsilon/F$ ($x10^{-4}$) | GaAs - $\Delta\varepsilon/F$ ($x10^{-5}$) | Transfer efficiency - % |
|---|---|---|---|
| Gold bonding | -1.329 ±0.056 | -3.360 ±0.28 | 25.28% ± 3.19% |
| SU8 bonding | -0.950 ±0.081 | -6.583 ±0.16 | 69.24% ± 7.83% |